\documentclass[aip, amsmath, amssymb, reprint, onecolumn]{revtex4-1}

\usepackage{amsthm}
\usepackage{amsfonts}
\usepackage{dsfont}
\usepackage{hyperref}
\usepackage{lipsum}
\usepackage{mathtools}

\usepackage[utf8]{inputenc}
\usepackage[T1]{fontenc}
\usepackage{mathptmx}
\usepackage{etoolbox}

\newtheorem{theorem}{Theorem}
\newtheorem{lemma}{Lemma}

\newtheorem{proposition}{Proposition}
\newtheorem{assumption}{Assumption}

\newcommand{\T}{\mathbb{T}}

\newcommand{\papertitle}{A simple counterexample to the Hohenberg--Kohn theorem for excited states} 
\newcommand{\paperabstract}{%
  For an even number of non-interacting fermions on the one-dimensional unit torus, a semi-explicit counterexample to the Hohenberg--Kohn theorem for excited states is exhibited.
  Indeed, for each $n = 2l$, it is shown that there exist constants $m \in (0, 1)$ and $\ell > 0$ such that an $n$-body system with the rescaled Lam\'e potential $v_*(x) = 2m \ell^2 \sn^2(\ell x; m)$ has a simple first excited energy and a first excited-state particle density which is constant on the torus.
  By translating this potential, we exhibit a non-trivial one-parameter family of potentials, all of which have a constant first excited-state density.
  As such, the potential-to-density map is not one-to-one in this setting.
  Under a spectral nonresonance assumption, we obtain the same nonuniqueness at a simple excited energy for odd $n \geq 3$.
  We also explore connections to the classification problem of elastica in the plane when $n = 2$, and outline some related questions of interest. 
}

\DeclareMathOperator{\sn}{sn}
\DeclareMathOperator{\cn}{cn}
\DeclareMathOperator{\dn}{dn}

\newcommand{\dd}{\mathrm{d}}

\DeclareMathOperator{\im}{Im}

\DeclarePairedDelimiter{\abs}{\lvert}{\rvert}

\DeclarePairedDelimiter{\inner}{\langle}{\rangle}

\DeclarePairedDelimiterX{\ketbra}[1]{|}{|}{#1\delimsize\rangle\delimsize\langle#1}

\makeatletter
\def\@email#1#2{%
  \endgroup
  \patchcmd{\titleblock@produce}
    {\frontmatter@RRAPformat}
    {%
      \frontmatter@RRAPformat{%
        \produce@RRAP{*#1\href{mailto:#2}{#2}}%
      }%
      \frontmatter@RRAPformat
    }
    {}%
    {}%
}
\makeatother

\begin{document}
  \title[\papertitle]{\papertitle}
  \author{D. Shillingford}
  \email{d.shillingford@mail.utoronto.ca}
  \affiliation{Department of Mathematics, University of Toronto}
  \date{\today}

  \begin{abstract}
    \paperabstract
  \end{abstract}

  \maketitle

  \section{Introduction}

  In quantum physics and quantum chemistry, density functional theory is a powerful and commonly used tool to predict the physical properties of many-body systems.
  The cornerstone of this theory is the Hohenberg--Kohn theorem, which informally states that the external potential applied to a many-body system of interacting quantum particles can be uniquely determined (up to an additive constant) from the particle density of the ground state of the system.
  This is not obvious as one cannot in general recover details about particle-particle correlations from the particle density of an arbitrary quantum system.
  Nonetheless, for the ground state, these correlations can be recovered by reverse engineering the external potential from the ground state density and recomputing the ground state wavefunction from this potential.

  With the success of the Hohenberg--Kohn theorem for ground states, one may ask if the same result is true for excited states. 
  This question was examined in detail in the work of Gaudoin and Burke\cite{GaudoinBurke}, who gave strong numerical evidence for the existence of two radial potentials that yield the same excited-state density for the configuration $1s^1 2s^2$, and two external potentials that yield the same first excited-state density for a pair of fermions in a box.
  In both cases, the particles are non-interacting.
  This is strong numerical evidence that the Hohenberg--Kohn theorem fails for excited states in general, for excited states with a fixed orbital configuration, and for excited states with a fixed excitation number.
  They also describe a mechanism for the latter failure by examining the susceptibility operator $\chi_k$ associated to a $k$-th excited-state density.
  
  The lack of a Hohenberg--Kohn theorem for non-interacting excited states was further examined by Li et al.\cite{Li}.
  For any energy eigenstate of two non-interacting fermions under a fixed external potential in one dimension, they derive a
  nonlinear differential equation whose non-trivial solutions, subject to appropriate boundary conditions, yield additional external potentials with energy eigenstates having the same density as the original state. 
  
  Li et al. further give strong numerical evidence for the existence of potentials for which non-trivial solutions to this equation exist.
  In contrast to the investigation of Gaudoin and Burke, the additional states generated by this method do not in general have the same excitation number or orbital configuration as the original state.
  Nonetheless, this provides additional numerical evidence for the failure of the Hohenberg--Kohn theorem for excited states in general. 

  While these numerical results leave little doubt that a general Hohenberg--Kohn theorem does not hold for excited states, neither work provides a rigorous existence proof for its counterexamples.
  Gaudoin and Burke construct their examples numerically and describe a mechanism for nonuniqueness, while Li et al. derive an exact differential equation but do not rigorously establish the existence of non-trivial solutions satisfying the required boundary and normalization conditions.
  In this letter, for any even number of particles, we give a semi-explicit construction of a non-constant potential on the unit torus that yields a constant first excited-state density.
  Translating this potential produces a one-parameter family of distinct potentials with the same first excited-state density, thereby proving the failure of a general Hohenberg--Kohn theorem for systems with an even number of particles and periodic boundary conditions, even at a fixed excitation number.
  Under a spectral nonresonance assumption, we obtain the same nonuniqueness at a simple excited level for odd $n \geq 3$. 
  
  \section{Main Results and Proofs}

  Before stating the results precisely, we introduce some notation.
  Let $\mathbb{T} = \mathbb{R} / \mathbb{Z}$ denote the (unit) torus, 
  \begin{equation}
    \mathcal{H} = \bigwedge^{n} L^2(\mathbb{T})
  \end{equation}
  be the state space of $n$ fermions on the torus, and for each $v \in C^\infty(\T)$ let $\Lambda_k(v)$ be the $k$-th eigenvalue of
  \begin{equation}
    \mathfrak{H}(v) = \operatorname{d\Gamma}{(-\Delta + v(\hat x))} = \sum_{k = 1}^{n} \left( -\Delta_k + v(\hat x_k) \right)
  \end{equation}
  in non-decreasing order, indexed from zero, with multiplicity.
  
  Let $\mathcal{V}_k = \{ [v] \in \dot C^\infty(\T) : \Lambda_k(v)\ \mathrm{is}\ \mathrm{simple} \}$ where $\dot C^\infty(\T)$ is the space of smooth functions on the torus modulo constant shifts
  and note that $\mathcal{V}_k$ is an open set by standard spectral perturbation theory.

  Whenever $[v] \in \mathcal{V}_k$, let $\eta_k(v) \in \mathcal{B}(\mathcal{H})$ be the projection onto the (one-dimensional) eigenspace of $\Lambda_k(v)$ 
  and $\rho_k(v) \in L^1(\T)$ the corresponding particle density satisfying
  \begin{equation}
    \int \mu(x) \rho_k(x; v) \, \dd x = \operatorname{Tr}{\left( \dd \Gamma(\mu(\hat x)) \eta_k(v) \right)} \, .  
  \end{equation}
  In particular, if $\eta_k(v) = |\phi_1 \wedge \ldots \wedge \phi_{n}\rangle\langle\phi_1 \wedge \ldots \wedge \phi_{n}|$ for some orthonormal $\phi_1, \ldots, \phi_n \in L^2(\T)$, we have $\rho_k(x; v) = \abs{\phi_1(x)}^2 + \ldots + \abs{\phi_{n}(x)}^2$.
  Our main result is the following.

  \begin{theorem}
    \label{maincoroll}

    Suppose $n$ is even.
    Then there exists $[v_*] \in \mathcal{V}_1$ such that the map $[v] \in \mathcal{V}_1 \mapsto \rho_1(v) \in L^1(\T)$ is not injective in any open neighbourhood of $[v_*]$.

    In particular, there exists a potential $\tilde v_{*} \in C^\infty(\T)$ such that $\rho_1(v_*) = \rho_1(\tilde v_{*})$ while $\tilde v_{*} - v_*$ is non-constant.
  \end{theorem}

  \noindent \emph{Remarks.}
  Although Theorem \ref{maincoroll} assumes that $n$ is even, we later state and prove a complementary but conditional result for odd $n \geq 3$, cf. Theorem \ref{maincoroll-odd}.
  In both Theorem \ref{maincoroll} and Theorem \ref{maincoroll-odd} the potential has the form $v_*(x) = 2 m \ell^2 \sn^2(\ell x; m)$, where $\sn$ is the elliptic sine function and the constants $m \in (0, 1)$ and $\ell > 0$ depend implicitly on $n$.
  Theorem \ref{maincoroll} will follow straightforwardly from the proposition below whose proof shows the existence of $m$ and $\ell$ such that $v_*$ has an excited state with a constant density. 
  
  The key ingredients in the proof of Proposition \ref{mainthm} are Lemma \ref{mainlemma} which shows that the squared moduli of Floquet eigenfunctions of $v_*$ relevant to the proof are affine in $x \mapsto \sn^2(\ell x; m)$
  and Lemma \ref{estimates} which gives endpoint behaviour of the normalized coefficients as they vary with the Jacobi parameter $m$. 

  \begin{proposition}
    \label{mainthm}
    Suppose $n$ is even. Then there exists a non-constant potential $v_* \in C^\infty(\T)$ such that 
    \begin{enumerate}
      \item $\Lambda_1(v_*)$, the first excited-state energy, is simple, and 
      \item $\rho_1(v_*)$ is constant, in particular, $\rho_1(x; v_*) = n$.  
    \end{enumerate}
  \end{proposition}

  \begin{proof}
    Let $v_*(x) = 2m\ell^2 \sn^2(\ell x; m)$
    for some $m \in (0, 1)$ and $\ell > 0$, depending on $n$, to be specified later, where $\sn$ is the elliptic sine function.
    
    It is well-known that for $m \in (0, 1)$, the square of the elliptic sine function is periodic with a period of $\ell_0(m) = 2 K\left(m\right)$.
    Here $K(m)$ is the complete elliptic integral of the first kind evaluated with the Jacobi parameter $m$.
    The resulting band structure of the Lam\'e Hamiltonian (of degree one) $H(m) = -\partial_x^2 + 2 m \sn^2(\hat x; m)$ is well-studied.

    Indeed, the band structure consists of one finite band $[\lambda_0, \lambda_{n - 1}]$ and a semi-infinite band $[\lambda_{n}, \infty)$, separated by a gap.
    The Floquet solutions $\phi_0, \phi_{n - 1}$ and $\phi_{n}$ corresponding to the band edges are periodic, anti-periodic, and anti-periodic respectively.
    These are the degree one Lam\'e polynomials\cite{Maier} and are proportional to $\dn(-; m)$, $\cn(-; m)$ and $\sn(-; m)$ respectively with $\lambda_0 = m$, $\lambda_{n - 1} = 1$ and $\lambda_{n} = 1 + m$, cf. \onlinecite[p. 205]{Arscott}.
    Here $\cn$ and $\dn$ are the elliptic cosine and delta amplitude functions respectively.

    Let $n = 2 l$.
    From Floquet theory, for each $k \in \mathbb{N}$ with $0 < k < l$, there exist further Floquet solutions $\phi_{2k - 1}$ and $\phi_{2k}$ associated to spectral points $\lambda_{2k - 1} = \lambda_{2k}$ within the band $[\lambda_0, \lambda_{n - 1}]$ satisfying  
    $\phi_{2k - 1}(x + \ell_0(m)) = e^{i \pi k / l} \phi_{2k - 1}(x)$ and $\phi_{2k}(x + \ell_0(m)) = e^{-i\pi k / l} \phi_{2k}(x)$.
    These points satisfy $\lambda_{2k - 2} < \lambda_{2k - 1} = \lambda_{2k} < \lambda_{2k + 1}$.
    It then follows that when $H(m)$ acts on the torus of period $\ell = 2 l \ell_0(m) = n \ell_0(m)$, 
    $\phi_0$ is the ground state while $\phi_1, \ldots, \phi_{n}$ are the first $n$ excited states.

    Using standard facts about fermionic many-body states, $\Lambda_0(v_*) = \ell^2 \sum_{k = 0}^{n - 1} \lambda_k$, $\Lambda_1(v_*) = \ell^2 \left( \lambda_{n} + \sum_{k = 0}^{n - 2} \lambda_k \right)$ and
    \[ \Lambda_2(v_*) = \ell^2 \times \min \left\{ \lambda_{n + 1} + \sum_{k = 0}^{n - 2} \lambda_k, \lambda_{n} + \lambda_{n - 1} + \sum_{k = 0}^{n - 3} \lambda_k \right\} \, . \] 
    
    We then have that $\Lambda_1(v_*)$ is simple if $\lambda_{n} - \lambda_{n - 1} > 0$, $\lambda_{n + 1} - \lambda_{n} > 0$ and $\lambda_{n - 1} - \lambda_{n - 2} > 0$.
    Indeed, the first inequality is true because of the band gap, while the other two are true by general Floquet theory.
    With
    \[ \phi_{k, \ell}(x) = \sqrt{\frac \ell n} \times \phi_k(\ell x) = \sqrt{\ell_0(m)} \times \phi_k(\ell x) \]
    it then follows by rescaling that $\eta_0(v_*) = |\phi_{0, \ell} \wedge \ldots \wedge \phi_{n - 1, \ell}\rangle \langle\phi_{0, \ell} \wedge \ldots \wedge \phi_{n - 1, \ell} |$ and
    \[ \eta_1(v_*) = \ketbra{\phi_{0, \ell} \wedge \ldots \wedge \phi_{n - 2, \ell} \wedge \phi_{n, \ell}} \, . \]

    We now calculate $\rho_1(v_*)$.
    To this end, it follows from Lemma \ref{mainlemma} that $\abs{\phi_k(x)}^2 = a_{\lambda_k}(m) - m b_{\lambda_k}(m) \sn^2(x; m)$. 
    We therefore calculate
    \[ \rho_1(x; v_*) = \ell_0(m) \left[ a_{\lambda_n}(m) - m b_{\lambda_n}(m) \sn^2(\ell x; m) + \sum_{k = 0}^{n - 2} \left( a_{\lambda_k}(m) - m b_{\lambda_k}(m) \sn^2(\ell x; m) \right) \right] \, . \]
    To obtain a constant density, it then suffices to find $m \in (0, 1)$ such that
    \[ f(m) = b_{\lambda_n(m)}(m) + \sum_{k = 0}^{n - 2} b_{\lambda_k(m)}(m) = 0 \, , \]
    where we emphasize the $m$-dependence of $\lambda_k$.
   
    Since $\lambda_n(m) = 1 + m$, $b_{\lambda_n(m)}(m) = c_1(m)$ (with $c_\zeta(-)$ as defined in Lemma \ref{estimates}) and thus
    \[ f(m) = c_1(m) + \sum_{k = 0}^{n - 2} b_{\lambda_k(m)}(m) \, . \]
    Indeed, by Lemma \ref{estimates}, $\lim_{m \to 1^-} c_1(m) = 0$ while $\lambda_k(m) \leq 1 \implies \lim_{m \to 1^-} b_{\lambda_k(m)}(m) = \frac{1}{2}$.
    We conclude that \[ \lim_{m \to 1^-} f(m) \geq \frac{n - 1}{2} \geq \frac{1}{2} \, . \]
   
    On the other hand, since $\lambda_k(m) \leq \lambda_{n - 2}(m) \leq 1$ and $b_\lambda(m)$ is an increasing function of $\lambda$ for fixed $m$ when $\lambda \leq 1$ we have
    \[ f(m) \leq c_1(m) + (n - 1) \times b_{\lambda_{n - 2}(m)}(m) \, . \]
    The potential $v_*(x) = 2m \ell^2 \sn^2(\ell x; m)$ converges uniformly to $0$ as $m \to 0$ and by spectral perturbation theory, it follows that $\lambda_{n - 2}(-)$ is continuous for sufficiently small $m$ and $\lambda_{n - 2}(m) = \lambda_{n - 2}(0) + o(1)$
    where
    \[ \lambda_{n - 2}(0) = \left( \frac{\pi (n - 2)}{\ell} \right)^2 = \left( \frac{\pi (n - 2)}{\pi n} \right)^2 = \left( 1 - \frac{2}{n} \right)^2 \, . \]
    Since $\lambda_{n - 2}(0)$ is strictly less than one, $\lambda_{n - 2}(m)$ is uniformly bounded away from one for sufficiently small $m$.
    It follows that
    \[ b_{\lambda_{n - 2}(m)}(m) \xrightarrow{m \to 0^+} \frac{1}{\pi} \times \frac{1}{1 - \lambda_{n - 2}(0)} \, . \] 
    This is bounded, while $c_1(m) \to -\infty$ as $m \to 0^+$ and therefore $\lim_{m \to 0^+} f(m) = -\infty$.

    Since $f$ is continuous on $(0, 1)$ by spectral perturbation theory, it follows by the intermediate value theorem that $f(m) = 0$ for some $m \in (0, 1)$.
    For this $m$, we have $\rho_1(v_*)$ constant, completing the proof.
  \end{proof}

  Theorem \ref{maincoroll} then follows as a corollary.

  \begin{proof}[Proof of Theorem \ref{maincoroll}]
    Let $v_*(x) = 2 m \ell^2 \sn^2(\ell x; m)$ for $m \in (0, 1)$ and $\ell > 0$ as in the proof of Proposition \ref{mainthm}
    and suppose for contradiction that $[v] \mapsto \rho_1(v)$ is injective in some neighbourhood of $[v_*]$.
    Let $v_{s}(x) = v_*(x + s)$.
    Since $v_*$ is not constant, the map $s \mapsto [v_s]$ is injective in some neighbourhood of $0$ and therefore $s \mapsto \rho_1(v_s)$ is injective in some neighbourhood of $0$.
    This, however, is a contradiction as $\rho_1(x; v_s) = n$ for all $s \in \T$.
  \end{proof}

  We now proceed to the statements and proofs of Lemma \ref{mainlemma} and Lemma \ref{estimates}.

  \begin{lemma}
    \label{mainlemma}

    Let $m \in (0, 1)$.
    Suppose $\phi$ satisfies the Lam\'e equation
    \[ -\phi''(x) + 2 m \sn^2(x; m) \phi(x) = \lambda \phi(x) \]
    such that $\phi$ is quasiperiodic with $\phi(x + \ell_0(m)) = \xi \phi(x)$ for some $\abs{\xi} = 1$
    and satisfies the normalization condition
    \[ \int_0^{\ell_0(m)} \abs{\phi(x)}^2 \, \dd x = 1 \, . \]
    Suppose further that either $\im \xi \neq 0$ or $\lambda \in \{ m, 1, 1 + m \}$.

    Then $\abs{\phi(x)}^2 = a_\lambda(m) - m b_\lambda(m) \sn^2(x; m)$
    where
    \[ a_\lambda(m) = \frac{1 + m - \lambda}{(m - \lambda) \ell_0(m) + \tilde \ell_0(m)}\, , \quad  b_\lambda(m) = \frac{1}{\tilde \ell_0(m) - (\lambda - m) \ell_0(m)} \, , \]
    and $\tilde \ell_0(m) = 2 E(m)$.
  \end{lemma}

  \noindent \emph{Remark.} A similar construction in which the density of a state is an affine function of $\sn^2$ can be found in the work of Thies\cite{Thies}.

  \begin{proof}
    We consider two cases, $\lambda \in \{ m, 1, 1 + m \}$ and $\im \xi \neq 0$.
    The former case concerns the three band edges of the Lam\'e Hamiltonian which are explicitly known.
    When $\lambda = m$, there is a unique solution (up to normalization) $\dn(x; m)$ with $\xi = 1$.
    When $\lambda = 1$ or $\lambda = 1 + m$, the unique solutions are $\cn(x; m)$ and $\sn(x; m)$ respectively with $\xi = -1$.
    In these explicit cases, we have respectively
    \[ \abs{\phi(x)}^2 \propto \dn^2(x; m) = 1 - m \sn^2(x; m); \quad \abs{\phi(x)}^2 \propto \cn^2(x; m) = 1 - \sn^2(x; m);\ \mathrm{and} \quad \abs{\phi(x)}^2 \propto \sn^2(x; m) \, . \] 
    Using that
    \[ \int_0^{\ell_0(m)} \sn^2(x; m) \, \dd x = \frac{1}{m} \left( \ell_0(m) - \tilde \ell_0(m) \right) \, , \]
    we conclude that in the three cases above, we have
    \[ \abs{\phi(x)}^2 = \frac{1 - m \sn^2(x; m)}{\tilde\ell_0(m)}; \quad \abs{\phi(x)}^2 = \frac{m - m\sn^2(x; m)}{ \tilde \ell_0(m) - (1 - m) \ell_0(m)};\ \mathrm{and} \quad \abs{\phi(x)}^2 = \frac{m\sn^2(x; m)}{\ell_0(m) - \tilde \ell_0(m)}  \, . \] 
    
    It remains to consider the case $\im \xi \neq 0$.
    To this end, one can show that if $\phi$ and $\tilde \phi$ are two solutions to the Lam\'e equation with the same eigenvalue then their product
    $u = \phi \tilde \phi$ satisfies the third-order equation
    \begin{equation}
      \label{threeorder} u'''(x) + 4 (\lambda - 2 m \sn^2(x; m)) \, u'(x) - 4 m \left[ \sn^2(x; m) \right]' u(x) = 0 {\ }\cite{Arscott, HaeseHill}. 
    \end{equation}

    Since $\im \xi \neq 0$, $\phi$ and $\bar \phi$ are two linearly independent solutions to the Lam\'e equation.
    It follows that $\phi^2$, $\bar\phi^2$, and $\abs{\phi}^2$ are three solutions to \eqref{threeorder}.
    Additionally, these solutions are linearly independent by the Wronskian identity
    \[ \operatorname{W}{\left(\phi^2, \abs{\phi}^2, \bar\phi^2\right)} = 2 \times \operatorname{W}{(\phi, \bar \phi)}^3 {\ } \cite{Seminar}. \]
    Since $\im \xi \neq 0$, only $\abs{\phi}^2$ is $\ell_0(m)$-periodic and thus any other periodic solution to \eqref{threeorder} must be proportional to $\abs{\phi}^2$. 

    By direct computation, one can verify that
    \begin{align}
      \left[ \sn^2(x; m) \right]''' 
      &= [ 12 m \sn^2(x; m) - 4 (m + 1) ] \times \left[ \sn^2(x; m) \right]'
    \end{align}
    and therefore by matching coefficients,
    $u(x) = c - \sn^2(x; m)$ is a periodic solution to \eqref{threeorder} if
    \[ 4(m + 1) - 4 \lambda - 4m c = 0 \implies c = \frac{m + 1 - \lambda}{m} \, . \]
   
    By the normalization condition,
    \[ \abs{\phi(x)}^2 = \frac{c - \sn^2(x; m)}{c \ell_0(m) - \frac{1}{m} \left( \ell_0(m) - \tilde \ell_0(m) \right)} = \frac{m + 1 - \lambda - m \sn^2(x; m)}{(m - \lambda) \ell_0(m) + \tilde \ell_0(m)} . \]
  \end{proof}  

  \begin{lemma}
    \label{estimates}
    Let $b_\lambda : (0, 1) \to \mathbb{R}$ be defined as in Lemma \ref{mainlemma}
    and let $c_\zeta : (0, 1) \to \mathbb{R}$ such that
    \[ c_\zeta(m) = \frac{1}{\tilde \ell_0(m) - \zeta \ell_0(m)} = b_{\zeta + m}(m) \, . \]
 
    Then the following hold:
    \begin{enumerate}
      \item $\lim_{m \to 0^+} c_1(m) = -\infty$
      \item $\lim_{m \to 0^+} b_\lambda(m) = \frac{1}{\pi} \times \frac{1}{1 - \lambda}$ uniformly in $\lambda$ for $\lambda$ uniformly bounded away from 1 and $\infty$.
      \item $\lim_{m \to 1^-} b_\lambda(m) = \frac{1}{2}$ uniformly in $\lambda$ for $m \leq \lambda \leq 1$, i.e., $\lim_{m \to 1^-} \left( \sup_{m \leq \lambda \leq 1} \abs*{b_\lambda(m) - \frac{1}{2}} \right) = 0$
      \item $\lim_{m \to 1^-} c_\zeta(m) = 0$ uniformly in $\zeta$ for $\zeta \geq 1$. 
    \end{enumerate}
  \end{lemma}

  \begin{proof}
    To begin, we have
    \[ c_1(m) = \frac{1}{2} \times \frac{1}{E(m) - K(m)} \, . \]
    $\lim_{m \to 0^+} c_1(m) = -\infty$ as $E(0) = K(0) = \frac{\pi}{2}$ while $K(m) \geq E(m)$ for all $m \in (0, 1)$.

    Next, we show the desired uniform limit for $b_\lambda(m)$ as $m \to 0^+$.
    In this case, $E(m) - (\lambda - m) K(m)$ converges uniformly to $\frac{\pi}{2} - \frac{\pi}{2} \times \lambda$ on every bounded set.
    Since $\lambda$ is uniformly bounded away from one, $\frac{\pi}{2} - \frac{\pi}{2} \times \lambda$ is uniformly bounded away from zero giving the desired result.

    We now proceed to the uniform convergence result for $m \leq \lambda \leq 1$.
    For all $m \leq \lambda \leq 1$, we have
    \[ E(m) \geq E(m) - (\lambda - m) K(m) \geq E(m) - (1 - m) K(m) \, . \]
    $E(m) - (1 - m) K(m) \geq 0$ as 
    \[ \int_0^\frac{\pi}{2} \left( 1 - m \sin^2(t) \right)^{1 / 2} \, \dd t - (1 - m) \int_0^\frac{\pi}{2} \left( 1 - m \sin^2(t) \right)^{-1 / 2} \, \dd t = \int_0^\frac{\pi}{2} \frac{m \cos^2(t)}{(1 - m \sin^2(t))^{1 / 2}} \, \dd t \geq 0 \]
    and therefore
    \[ \frac{1}{E(m)} \leq \frac{1}{E(m) - (\lambda - m) K(m)} = 2 b_\lambda(m) \leq \frac{1}{E(m) - (1 - m) K(m)} \, . \]
    By the squeeze theorem, $2 b_\lambda(m) \to E(1)^{-1} = 1$ uniformly for $m \leq \lambda \leq 1$ as $m \to 1^-$ giving the result. 

    To conclude, for $\zeta \geq 1$ we have $\zeta K(m) - E(m) \geq K(m) - E(m) > 0$ (for $m \in (0, 1)$) and thus
    \[ \abs{2 c_\zeta(m)} = \frac{1}{\zeta K(m) - E(m)} \leq \frac{1}{K(m) - E(m)} \xrightarrow{m \to 1^-} 0 \,  . \]
  \end{proof}

  To prove the version for an odd number of particles, we need some extra notation.

  \begin{lemma}
    \label{mmm}
    Let $H(m) = -\partial_x^2 + 2 m \sn^2(\hat x; m)$.
    Then for each $r \in \mathbb{R}$ with $r \geq 0$ there exists a continuous function $\lambda_r : (0, 1) \to \mathbb{R}$
    with $\lambda_r(m) \to r^2$ as $m \to 0^+$
    such that for every $m \in (0, 1)$ there exists $\psi_{r, m}$ satisfying the following:
    \begin{enumerate}
      \item $\psi_{r, m}$ and $\bar \psi_{r, m}$ are Floquet eigenfunctions of $H(m)$ with unit norm over one quasiperiod $\ell_0(m)$ and eigenvalue $\lambda_r(m)$
      \item $\psi_{r, m}(x + \ell_0(m)) = e^{i\pi r} \psi_{r, m}(x)$
      \item $\psi_{r, m}$ and $\bar \psi_{r, m}$ are orthogonal over an interval of $d \ell_0(m)$ whenever $d \in \mathbb{N}$, and $r \not\in \mathbb{Z}$ while $dr \in \mathbb{Z}$
    \end{enumerate}

    Furthermore, $m \leq \lambda_r(m) \leq 1$ if $0 < r < 1$ and $\lambda_r(m) \geq 1 + m$ if $r > 1$. 
  \end{lemma}

  \noindent \emph{Remark.} As in the proof of Proposition \ref{mainthm}, we have seen that since $H(0)$ is the free Hamiltonian, there are eigenvalues of $H(m)$ close to those of the free Hamiltonian for small $m \in (0, 1)$.
  This lemma goes a step further and shows that after fixing the wave~number $r$, we can track how the corresponding eigenvalue $\lambda_r(m)$ evolves from $\lambda_r(0) = r^2$ for all $m \in (0, 1)$ in a continuous way.

  \begin{proof}
    To fix a common domain, consider the rescaled Hamiltonian
    \[ \tilde H(m) = -\partial_y^2 + 2 m \ell_0(m)^2 \sn^2(\ell_0(m) \, \hat y; m) \]
    which is a periodic Hamiltonian with unit period.
    This Hamiltonian is unitarily equivalent to $H(m)$ up to a factor of $\ell_0(m)^2$
    and therefore has the same spectrum after rescaling.

    Let $\tilde H(m)$ act on the `twisted' unit torus, i.e., on an appropriate extension of
    $\{ \psi \in C^\infty(\mathbb{R}) : \psi(x + 1) = e^{i \pi r} \psi(x) \}$
    with the standard inner product over one quasiperiod.
    
    $\pi^2 r^2$ is an eigenvalue of $\tilde H(0)$ with, e.g., $\tilde \psi_{r, 0}(y) \propto e^{i \pi r y}$.
    Since $m \in [0, 1) \mapsto 2 m \ell_0(m)^2 \sn^2(\ell_0(m) \, \hat y ; m)$ is continuous in the operator norm, standard spectral perturbation theory implies that there exists a continuous function $\tilde \lambda_r(m)$ so that $\tilde \lambda_r(m)$ is an eigenvalue of $\tilde H(m)$ with $\tilde \lambda_r(m) \to \pi^2 r^2$ as $m \to 0$.
    We claim that
    \[ \lambda_r(m) = \frac{\tilde \lambda_r(m)}{\ell_0(m)^2} \]
    satisfies the conditions of the lemma.

    Since $\ell_0$ is continuous on $(0, 1)$, $\lambda_r$ is continuous on $(0, 1)$ with
    \[ \lim_{m \to 0^+} \lambda_r(m) = \frac{\pi^2 r^2}{\ell_0(0)^2} = r^2 \, . \]

    Fix $m \in (0, 1)$.
    By the rescaling, $\lambda_r(m)$ is an eigenvalue of $H(m)$.
    Let $\tilde \psi_{r, m}$ be any normalized eigenfunction of $\tilde H(m)$ with eigenvalue $\tilde \lambda_r(m)$.
    It then follows from the boundary condition that $\tilde \psi_{r, m}(y + 1) = e^{i \pi r} \tilde \psi_{r, m}(y)$.
    Additionally, if $d \in \mathbb{N}$ is such that $r$ is not an integer while $d r$ is an integer,
    \[ \inner{\bar{\tilde \psi}_{r, m}, \tilde \psi_{r, m}} = \int_0^{d} \tilde \psi_{r, m}^2(y) \, \dd y = \sum_{k = 0}^{d - 1} e^{2i \pi k r} \int_0^1 \tilde \psi_{r, m}^2(y) \, \dd y = \frac{1 - e^{2i \pi dr}}{1 - e^{2i \pi r}} \int_0^1 \tilde \psi_{r, m}^2(y) \, \dd y = 0 \, . \]
    Rescaling, $\psi_{r, m}(x) \propto \tilde \psi_{r, m}(x / \ell_0(m))$ has the desired properties of the lemma after normalization. 

    To conclude, note that by the band theory of the Lam\'e Hamiltonian, we must have $m \leq \lambda_r(m) \leq 1$ or $\lambda_r(m) \geq 1 + m$
    for each $m \in (0, 1)$.
    Fix $0 < r < 1$.
    To show that $0 \leq \lambda_r(m) \leq 1$ for all $m \in (0, 1)$ it therefore suffices to show that $\lambda_r(m) < 1 + m$ for all $m \in (0, 1)$.
    We proceed by contradiction.
    Suppose that $\lambda_r(m_1) \geq 1 + m_1$ for some $m_1 \in (0, 1)$.
    Since $\lambda_r$ is continuous with $\lambda_r(m) \to r^2 < 1$ as $m \to 0^+$ there exists $m_0 \in (0, 1)$ so that $\lambda_r(m) < 1$ for all $0 < m \leq m_0$.
    It follows that $m_1 > m_0$.

    Since $\lambda_r(m_0) < 1$ we have $\lambda_r(m_0) - \frac{m_0}{2} < 1$.
    On the other hand, $\lambda_r(m_1) \geq 1 + m_1 \implies \lambda_r(m_1) - \frac{m_1}{2} \geq 1 + \frac{m_1}{2} > 1$.
    By the intermediate value theorem, there exists $m^* \in (m_0, m_1)$ such that
    \[ \lambda_r(m^*) = 1 + \frac{m^*}{2} \]
    but this contradicts the established fact that $m \leq \lambda_r(m) \leq 1$ or $\lambda_r(m) \geq 1 + m$ as $1 < 1 + \frac{m^*}{2} < 1 + m^*$.
    We therefore conclude that $0 \leq \lambda_r(m) \leq 1$.
    A similar proof establishes $\lambda_r(m) \geq 1 + m$ when $r > 1$.
  \end{proof}

  \begin{lemma}
    \label{rrr}
    Let $l \in \mathbb{N}$ and let $p_1, \ldots, p_l \in \mathbb{Q}$. 
    Suppose that $p_k > 1$ for all $1 \leq k \leq l$ and 
    \begin{equation}
    \label{rrr-est} \sum_{k = 1}^l \left( \frac{1}{p_k^2 - 1} \right) > \frac{1}{2} \, .
    \end{equation}

    Then there exists $m \in (0, 1)$ so that
    \begin{equation} \label{rrr-result}
      b_m(m) + 2 \sum_{k = 1}^l b_{\lambda_{p_k}(m)}(m) = 0
    \end{equation}
    where $\lambda_r$ is defined as in Lemma \ref{mmm} and $b_\lambda$ is defined as in Lemma \ref{mainlemma}.
  \end{lemma}

  \begin{proof}
    Since $p_k > 1$, it follows by Lemma \ref{mmm} that $\lambda_{p_k}(m) \geq 1 + m$.
    Let $\zeta_{p_k}(m) = \lambda_{p_k}(m) - m$ so that $\zeta_{p_k}(m) \geq 1$.
    It then follows that
    \[ b_m(m) + 2 \sum_{k = 1}^l b_{\lambda_{p_k}(m)}(m) = b_m(m) + 2 \sum_{k = 1}^l c_{\zeta_{p_k}(m)}(m) \, . \]

    By Lemma \ref{estimates}, as $m \to 1^-$, $b_m(m) \to \frac{1}{2}$, while since $\zeta_{p_k}(m) \geq 1$, $c_{\zeta_{p_k}(m)}(m) \to 0$.
    As a result
    \[ \lim_{m \to 1^-} \left( b_m(m) + 2 \sum_{k = 1}^l c_{\zeta_{p_k}(m)}(m) \right) = \frac 12 \, . \]
    
    On the other hand, for sufficiently small $m$, each $\lambda_{p_k}(m)$ is uniformly bounded away from one (as $p_k > 1$ and $\lambda_{p_k}$ is continuous)
    and since there are only finitely many, this can be made uniform in $k$ as well.
    Again, by Lemma \ref{estimates}, 
    \begin{equation}
      \label{rrr-m-zero} \lim_{m \to 0^+} \left( b_m(m) + 2 \sum_{k = 1}^l b_{\lambda_{p_k}(m)}(m) \right) = \frac{1}{\pi} + \frac{2}{\pi} \sum_{k = 1}^l \left( \frac{1}{1 - p_k^2} \right) \, .
    \end{equation}
    By \eqref{rrr-est}, the right-hand side of \eqref{rrr-m-zero} is negative, and therefore by the intermediate value theorem, there exists some $m \in (0, 1)$ such that 
    $b_m(m) + 2 \sum_{k = 1}^l b_{\lambda_{p_k}(m)}(m) = 0$.
  \end{proof}

  \begin{assumption} \label{nonresonance}
    For every $l \geq 1$, there exist distinct $p_1, \ldots, p_l \in \mathbb{Q} \setminus \mathbb{Z}$ satisfying the hypotheses of Lemma \ref{rrr}, such that, for at least one root $m \in (0,1)$ satisfying \eqref{rrr-result} guaranteed by that lemma, the energy
    \[ m + 2\sum_{k=1}^l \lambda_{p_k}(m) \]
    has a unique representation, up to permutation, as a sum of $2l + 1$ eigenvalues of $H(m)$ on $\mathbb{T}_{2d\ell_0(m)}$, with no eigenvalue used more often than its spectral multiplicity. Here $d$ is the least common denominator of the $p_k$.
  \end{assumption}

  \begin{theorem}
    \label{maincoroll-odd}

    Suppose $n$ is odd with $n \geq 3$ and Assumption \ref{nonresonance} holds.
    Then there exists an index $k \in \mathbb{N}$ and $[v_*] \in \mathcal{V}_k$ such that the map $[v] \in \mathcal{V}_k \mapsto \rho_k(v) \in L^1(\T)$ is not injective in any open neighbourhood of $[v_*]$.

    In particular, for this index $k$, there exists a potential $\tilde v_{*} \in C^\infty(\T)$ such that $\rho_k(v_*) = \rho_k(\tilde v_{*})$ while $\tilde v_{*} - v_*$ is non-constant.
  \end{theorem}

  \noindent \emph{Remark.}
  The proof of Theorem \ref{maincoroll-odd} is essentially identical to the proof of Theorem \ref{maincoroll} above after establishing Proposition \ref{mainthm-odd} below.
  We therefore refer the interested reader to the proof of Theorem \ref{maincoroll} above. 

  \begin{proposition}
    \label{mainthm-odd}
    Suppose $n$ is odd with $n \geq 3$ and Assumption \ref{nonresonance} holds.
    Then there exists a non-constant potential $v_* \in C^\infty(\T)$ and an index $k \in \mathbb{N}$ such that 
    \begin{enumerate}
      \item $\Lambda_k(v_*)$, the $k$-th excited-state energy, is simple, and 
      \item $\rho_k(v_*)$ is constant, in particular, $\rho_k(x; v_*) = n$.  
    \end{enumerate}
  \end{proposition}

  \begin{proof}
    Since $n$ is odd and greater than or equal to three, let $n = 2 l + 1$ with $l \in \mathbb{N}$.
    Let $p_1, \ldots, p_{l} \in \mathbb{Q} \setminus \mathbb{Z}$ be a set of non-integer rationals satisfying the hypotheses of Lemma \ref{rrr} that is non-resonant, whose existence is guaranteed by Assumption \ref{nonresonance},
    and let $m \in (0, 1)$ be the corresponding non-resonant Jacobi parameter for which
    \[ b_m(m) + 2 \sum_{k = 1}^{l} b_{\lambda_{p_k}(m)}(m) = 0 \, . \]
    Let $\ell = 2 d \ell_0(m)$ where $\ell_0$ is defined as in the proof of Proposition \ref{mainthm} and $d$ is the least common denominator of $\{ p_1, \ldots, p_{l} \}$. 
    As in the proof for even $n$, take $v_*(x) = 2m \ell^2 \sn^2(\ell x; m)$. 

    Let $\phi_0(x) \propto \dn(\ell x; m)$ be normalized and, for each $1 \leq k \leq l$, let $\phi_k(x) = \sqrt{\ell_0(m)} \times \psi_{p_k, m} (\ell x)$ where $\psi_{r, m}$ satisfies the following:
    \begin{enumerate}
      \item $\psi_{r, m}$ and $\bar \psi_{r, m}$ are Floquet eigenfunctions of $H(m)$ with unit norm over one quasiperiod $\ell_0(m)$ and eigenvalue $\lambda_r(m)$
      \item $\psi_{r, m}(x + \ell_0(m)) = e^{i\pi r} \psi_{r, m}(x)$
      \item $\psi_{r, m}$ and $\bar \psi_{r, m}$ are orthogonal over a period of $\ell$ whenever $r \not\in \mathbb{Z}$ while $dr \in \mathbb{Z}$
    \end{enumerate}
    with $\lambda_r(-)$ as defined in Lemma \ref{mmm}. Indeed, such a choice of $\psi_{r, m}$ exists by Lemma \ref{mmm}. 
    Lastly, let 
    \[ \Phi = \phi_0 \wedge \phi_1 \wedge \bar \phi_1 \wedge \ldots \wedge \phi_{l} \wedge \bar \phi_{l} \, . \]
    
    It then follows that $\Phi$ is an eigenfunction of $\mathfrak{H}(v_*)$ with eigenvalue
    \[ \Lambda = \ell^2 \left( m + 2 \sum_{k = 1}^{l} \lambda_{p_k}(m) \right) \, . \]
    In particular, $\Lambda = \Lambda_i(v_*)$ for some $i \in \mathbb{N}$, $\Lambda_i(v_*)$ is a simple eigenvalue by Assumption \ref{nonresonance} and $\eta_i(v_*) = \ketbra{\Phi}$.

    We then calculate $\rho_i(v_*)$.
    Indeed, similarly to the case of an even number of particles considered in Proposition \ref{mainthm},
    \[ \rho_i(x; v_*) = \ell_0(m) \left( a_m(m) - m b_m(m) \sn^2(\ell x; m) + 2 \sum_{k = 1}^{l} \left( a_{\lambda_{p_k}(m)}(m) - m b_{\lambda_{p_k}(m)}(m) \sn^2(\ell x; m) \right) \right) \]
    by Lemma \ref{mainlemma} and is therefore a constant function as 
    \[ b_m(m) + 2 \sum_{k = 1}^{l} b_{\lambda_{p_k}(m)}(m) = 0 \, . \]
  \end{proof}

  \section{Methodology and connections to previous work}

  The construction used in the proof of Proposition \ref{mainthm} was originally suggested by the large language model (LLM) GPT-5.
  The conditional extension to the case of an odd number of particles given in Proposition \ref{mainthm-odd} was later suggested by the LLM GPT-6.
  All details of the proof have been independently checked by the author.
  This letter is written entirely by the author aside from minor copy-editing performed by an LLM.
  While LLMs are excellent at proving theorems, they are notoriously bad at reconstructing the mathematical inspiration and precedent behind their constructions.
  Nonetheless, we attempt to identify some plausible mathematical antecedents to these constructions in this section.
  
  The key idea behind Theorems \ref{maincoroll} and \ref{maincoroll-odd} is to find a non-constant potential with an excited many-body state that has a constant particle density on the torus.
  By translating the potential, we create a one-parameter family of potentials that yield the same excited-state particle density.
  The spectrum of a Schr\"odinger operator on the torus is determined by a Hill equation and can be understood with Floquet theory. 
  This approach is natural and perhaps could have been identified by a clever (human) mathematician.
  On the other hand, identifying a particular periodic potential with a constant excited-state density requires an additional idea.
 
  \subsection{Connections to elastica}

  A possible route to the Lam\'e equation in particular is to consider the $n = 2$ case.
  In this case, we look for two real-valued functions $u$ and $w$ on the torus that are eigenfunctions of a Schr\"odinger operator with a common potential and satisfy 
  \begin{equation}
    \label{constraints}
    u^2(x) + w^2(x) = 2\, ,\int u(x) w(x) \, \dd x = 0 \, ,\  \mathrm{and}\ \int u^2(x) \, \dd x = \int w^2(x) \, \dd x = 1  \, .
  \end{equation}
  The set $\mathcal{S} = \{ (u, w) \in C^\infty(\T) \times C^\infty(\T) : \eqref{constraints} \ \mathrm{holds} \}$
  is closely related to an infinite-dimensional Stiefel manifold.
  It is natural in this setting to consider the constrained variational problem for the kinetic energy 
  \begin{equation}
    \label{energy} T(u, w) = \int  u'(x)^2 + w'(x)^2 \, \dd x
  \end{equation}
  on the set $\mathcal{S}$, since stationarity under these constraints yields Schr\"odinger equations.
  
  The Euler--Lagrange equations for this problem are
  \begin{equation} \label{eulerlagrange}
  -\begin{pmatrix} u(x) \\ w(x) \end{pmatrix}'' + v(x)\begin{pmatrix} u(x) \\ w(x) \end{pmatrix} = \begin{pmatrix} \lambda_u & \zeta \\ \zeta & \lambda_w \end{pmatrix} \begin{pmatrix} u(x) \\ w(x) \end{pmatrix} \, ,
  \end{equation}
  where $\lambda_u$, $\lambda_w$, $\zeta$ and $v$ are seen as Lagrange multipliers.
  Since the constraints are redundant, stationarity at $(u, w)$ will only uniquely determine $v$ up to an additive constant. 
  
  Suppose $(u, w)$ is a stationary point of \eqref{energy} over $\mathcal{S}$.
  The matrix on the right-hand side is symmetric and therefore decomposes as $R D R^{-1}$ with $D$ diagonal and $R \in \operatorname{SO}{(2)}$. 
  As such, after a rotation, we can find two functions $\tilde u$ and $\tilde w$ satisfying \eqref{constraints} so that
  \[ -\begin{pmatrix} \tilde u(x) \\ \tilde w(x) \end{pmatrix}'' + v(x)\begin{pmatrix} \tilde u(x) \\ \tilde w(x) \end{pmatrix} = \begin{pmatrix} \tilde \lambda_u & 0 \\ 0 & \tilde \lambda_w \end{pmatrix} \begin{pmatrix} \tilde u(x) \\ \tilde w(x) \end{pmatrix} \, . \]
  Therefore, any stationary point of \eqref{energy} over $\mathcal{S}$ yields a potential with a pair of orthonormal eigenfunctions whose squared amplitudes sum to a constant. 
  
  There is a close parallel with the C. Neumann system, an integrable system describing the motion of a classical particle confined to a sphere under the influence of a quadratic potential 
  \[ q \mapsto \frac{1}{2} \inner{q, Aq} \, . \]
  By the Neumann--Hill correspondence, a trajectory $q = (u,w)$ on the sphere of radius $\sqrt{2}$ gives rise to a Hill potential $v$ for which \eqref{eulerlagrange} holds with $A$ on the right-hand side \cite{Veselov}.
  As in our variational problem, $v$ is the Lagrange multiplier enforcing the pointwise constraint $u^2+w^2=2$.
  However, it is not clear how one could use this correspondence to find a stationary point of $T$.
  Indeed, one would still need to choose $A$ and find a periodic trajectory whose components satisfy the integral conditions in \eqref{constraints}.

  Instead, to show the existence of stationary points, there is a beautiful connection to the theory of elastica in the complex plane.
  Recall that each smooth function $\gamma \in \operatorname{Imm}{(\T; \mathbb{C})}$ parameterizes a smooth closed planar curve.
  The elastica problem is to find stationary points of the bending energy functional
  \[ B(\gamma) = \frac{1}{2} \int_0^1 \kappa_\gamma^2(t) \abs{\gamma'(t)} \, \dd t \, , \]
  subject to the length constraint
  \begin{equation} \label{length}
  \ell(\gamma) = \int_0^1 \abs{\gamma'(t)} \, \dd t = 1 \, .
  \end{equation}
  Here $\kappa_\gamma$ is the curvature of the curve and if $\gamma$ is a unit-speed parameterization, i.e., $\abs{\gamma'} = 1$, we have $\kappa_\gamma(t) = \abs{\gamma''(t)}$.

  The key connection to elastica is that if $(u, w) \in \mathcal{S}$ then
  \begin{equation}
  \label{uwrepresentation}
    \gamma_{u, w}(t) = \frac{1}{2} \int_0^t (u(s) + i w(s))^2 \, \dd s
  \end{equation}
  is a unit-speed immersion from $\T$ to $\mathbb{C}$ with $\ell(\gamma_{u, w}) = 1$\cite{Younes} and $B(\gamma_{u, w}) = T(u, w)$. 
  
  Additionally, the mapping $(u, w) \in \mathcal{S} \mapsto \gamma_{u, w} \in \operatorname{Imm}{(\T; \mathbb{C})}$
  is smooth (as a mapping between Fr\'{e}chet manifolds)
  and thus $T$ is stationary at $(u, w)$ whenever $B$ is stationary at $\gamma_{u, w}$ subject to \eqref{length}.

  The elastica problem is classical, and the stationary points of $B$ have been fully classified, cf. \onlinecite[Chap.~XIX]{Love}.
  In particular, among other solutions, it was known to Euler that $\gamma$ is a stationary point whenever $\gamma$ parameterizes the figure-eight elastica\cite{Euler}. 
  The figure-eight elastica, rescaled to unit length and parametrized by arclength, is of the form \(\gamma_{u,w}\) with
  \[ u(t) = \sqrt{2} \, \dn(\ell t;m) \, , \qquad w(t) = \sqrt{2m} \, \sn(\ell t;m) \, , \]
  where $\ell = 2 \ell_0(m)$ and $m \in (0,1)$ is the unique value satisfying $2E(m) = K(m)$,
  cf. \onlinecite{Sachkov, Miura}. 
  This parameterization can be recovered from the explicit form given in \onlinecite[Theorem~2.15, Case~II]{Miura} by recognizing
  \[ u(t) + i w(t) = \sqrt{2} \exp{\left( \frac{i \theta_w(\ell t)}{2} \right)} \]
  with
  \[ \theta_w(s) = 2 \arcsin \left( \sqrt m \sn(s; m) \right) \, . \]
  Once $u$ and $w$ are fixed, $v(t) = 2 m \ell^2 \sn^2(\ell t; m)$ follows from \eqref{eulerlagrange} up to an additive constant. 

  Unfortunately, the connection to elastica does not seem to generalize well beyond $n = 2$.
  Generalizations of \eqref{uwrepresentation} do exist for $n > 2$.
  Indeed, for $q(t) = (q_1(t), \ldots, q_n(t))$ with 
  \begin{equation}
    \label{constraints-d} \abs{q(x)}^2 = n, \ \mathrm{and}\ \int q(x) q(x)^\mathsf{T} \, \dd x = \mathds{1}
  \end{equation}
  generalizing \eqref{constraints},
  the map $q \mapsto \gamma_q \in \operatorname{Imm}{(\T; \mathbb{R}^{n \times n})}$ with $\gamma_q'(t) \propto q(t) q(t)^\mathsf{T} - \mathds{1}$
  yields a unit-speed parameterization of a curve whose bending energy is proportional to the obvious generalization of \eqref{energy}.
  An obstruction is that in $n = 2$, if we allow anti-periodic $(u, w)$ the map \eqref{uwrepresentation} can in fact represent \emph{any} smooth closed immersed planar curve (modulo translation and rescaling).
  It is therefore no surprise that the figure-eight elastica admits the representation above, as the only potential obstruction was that $(u, w)$ may have been anti-periodic.
  On the other hand, for $n > 2$, the condition that $\gamma'_q(t)$ is proportional to $q(t) q(t)^\mathsf{T} - \mathds{1}$ imposes a strong restriction which a higher-dimensional elastica need not satisfy. 

  \section{Discussion and Outlook}

  Theorem \ref{maincoroll} establishes the lack of a completely general Hohenberg--Kohn theorem for many-body excited states with an even number of particles while Theorem \ref{maincoroll-odd} establishes the corresponding conditional result for an odd number of particles.
  An obvious improvement would be to prove Theorem \ref{maincoroll-odd} unconditionally, perhaps by establishing that Assumption \ref{nonresonance} holds or through other means. 
  It is also of interest whether there is a fundamental difference between even and odd numbers of particles or if this is an artifact of the proof techniques used here.

  When $n$ is even, our unconditional result Theorem \ref{maincoroll} disproves the Hohenberg--Kohn theorem for non-interacting excited states on the one-dimensional torus at a fixed excitation level and the Hohenberg--Kohn theorem for non-interacting excited states on the one-dimensional torus with a fixed orbital configuration.
  Nonetheless, there may still exist domains, boundary conditions, or parameter regimes in which one (or both) of these theorems may hold true.
  The numerical evidence of Gaudoin and Burke strongly suggests the failure of the Hohenberg--Kohn theorem for non-interacting excited states in $(0, 1)$ with Dirichlet boundary conditions at a fixed excitation level, and in $\mathbb{R}^3$ with a fixed orbital configuration, cf. \onlinecite[Fig. 1 and Fig. 2]{GaudoinBurke}.
  In the latter case, following the spirit of this paper, one may be able to disprove the Hohenberg--Kohn theorem for non-interacting excited states on $\mathbb{R}^3$ for both a fixed excitation number and a fixed orbital configuration by constructing a potential $v_* : \mathbb{R}^3 \to \mathbb{R}$ that is not rotationally symmetric, but which has a rotationally symmetric excited-state density.

  While Theorem \ref{maincoroll} establishes the non-invertibility of $\rho_1$ at $v_*$, it is also of interest to understand the local mechanism for non-invertibility.
  In this direction, Gaudoin and Burke\cite{GaudoinBurke} investigated the G\^ateaux derivative of $\rho_k$, which is known as the linear susceptibility (of the $k$-th excited state) and denoted by $\chi_k$.
  Indeed, it follows by standard spectral perturbation theory that for fixed $v$, $\chi_k(v; -)$ is a linear integral operator. 
  Gaudoin and Burke\cite{GaudoinBurke} show that the number of non-trivial nonnegative eigenvalues of $\chi_1(v)$ (counted with multiplicity) is at most one.
  The constant functions contribute one trivial zero eigenvalue, counted with multiplicity.
  As such, the total number of nonnegative eigenvalues of $\chi_1(v)$ is at most two.
  For even $n$, Theorem \ref{maincoroll} implies that
  $ \partial_x v_* \in \operatorname{ker} \chi_1(v_*) $
  and therefore
  $ \operatorname{ker} \chi_1(v_*) = \operatorname{span}{\{ \mathbf{1}, \partial_x v_* \}}$.

  With an exact specification of the kernel of $\chi_1(v_*)$ there are two very natural questions of interest. 
  The first is whether the property 
  $\operatorname{dim} \operatorname{ker} \chi_1(v_*) = 2$
  is stable or unstable.
  In this direction, one may remain in the non-interacting setting by considering perturbations $v_* \to v_* + \epsilon \xi$ or introduce a weak two-body interaction, i.e.,
  \[ \mathfrak{H}(v) \to \mathfrak{H}(v) + \epsilon \dd \Gamma^{(2)}( w(\hat x, \hat y) ) \, . \]

  The second is whether a Lyapunov--Schmidt reduction is possible at $v_*$ which would show that all nearby potentials with constant first excited-state density are translates of $v_*$ up to an additive constant. 
  Such a reduction is possible if $\chi_1(v_*)$ is a Fredholm operator and $\chi_1$ is continuous in a neighbourhood of $v_*$.
  As topological properties, these are sensitive to the precise choice of function space and topology on which $\rho_1$ is posed.
  In this paper, we view $\rho_1$ as a map from an open subset of $\dot C^\infty(\T)$ to $L^1(\T)$ while Gaudoin and Burke implicitly view $\chi_1$ as a map from $L^2(\Omega)$ to $L^2(\Omega)$.
  Garrigue\cite{Garrigue} showed that $\rho_1$ is well-defined as a map from an open subset $\mathcal{V} \subseteq L^p(\Omega) + L^\infty(\Omega)$ into $W^{1, 1}(\Omega)$ for $p$ in some subset of $[1, \infty)$ depending on spatial dimension, that $\chi_1$ is continuous and that $\chi_1(v)$ is compact for all $v \in \mathcal{V}$. 
  In the latter instance, compactness is incompatible with Fredholmness and therefore no Lyapunov--Schmidt reduction is possible between these spaces.
  This raises the question of whether $\rho_1$ can be defined between suitable function spaces such that $\chi_1$ is continuous in a neighbourhood of $v_*$ and Fredholm at $v_*$. 
  In this direction, Corso and Laestadius\cite{Corso} apply the inverse function theorem to $\rho_0$ as a map from $H^{-1}(0, 1) / \mathbb{R}$ to the fixed-mass affine subspace of $H^1(0, 1)$ with Neumann boundary conditions by showing that $\chi_0$ is invertible (and therefore Fredholm). 
  
  \bibliography{paper}
\end{document}